\documentclass[preprint,authoryear,12pt]{elsarticle}

\usepackage{epsfig}
\usepackage{natbib}
\usepackage{amssymb,amsmath}
\usepackage{graphicx}
\usepackage{amssymb}

\usepackage[ps2pdf,%
a4paper=true,%
breaklinks=true,%
colorlinks=true,%
pdfauthor={First Author et al.},%
pdftitle={Template for manuscripts in Advances in Space Research}%
]{hyperref}

\journal{Advances in Space Research}

\begin{document}

%%%%%%%%%%%%%%%%%%%%%%%%%%%%%%%%%%%%%%%%%%%%%%%%%%%%%%%%%%%%%%%%%%%%%%%%%%%%%
%% Frontmatter
\begin{frontmatter}


% Use the tnoteref command within \title and fnref within \author or \address for footnotes;
% use the corref command within \author for corresponding author footnotes;
% \author{Name\corref{cor1}\fnref{label2}}
% \address{Address\fnref{label3}}

\title{Analysis of Ground Level Enhancements (GLE): Extreme solar energetic particle events have hard spectra}

\author[1]{E. Asvestari}
\author[2]{T. Willamo}
\author[1,3]{A. Gil}
\author[1,4]{I.G. Usoskin\corref{cor}}
\ead{ilya.usoskin@oulu.fi}
\author[5]{G.A. Kovaltsov}
\author[6]{V.V. Mikhailov}
\author[6]{A. Mayorov}

\address[1]{ReSoLVE Centre of Excellence, University of Oulu, FIN-90014 Oulu, Finland}
\address[2]{University of Helsinki, FIN-00014 Helsinki, Finland}
\address[3]{Institute of Mathematics and Physics, Siedlce University, Stanislawa Konarskiego 2, 08-110 Siedlce, Poland}
\address[4]{Sodankyl\"a Geophysical Observatory, University of Oulu, FIN-90014 Oulu, Finland}
\address[5]{Ioffe Physical-Technical Institute, Politechnicheskaya st. 26, 194021 St. Petersburg, Russia}
\address[6]{National research nuclear university "MEPhI", Kashirskoye sh. 31, 115409 Moscow, Russia}

\cortext[cor]{Corresponding author}

% Url can be given like this:
% \ead[url]{http://www.elsevier.com/wps/find/authorsview.authors/latex}

\begin{abstract}
Nearly 70 Ground Level Enhancements (GLEs) of cosmic rays have been recorded by the worldwide neutron monitor network since the 1950s
 depicting a big variety of energy spectra of solar energetic particles (SEP).
Here we studied a statistical relation between the event-integrated intensity of GLEs (calculated as count-rate relative
 excess, averaged over all available polar neutron monitors, and expressed in percent-hours) and the hardness of the solar particle energy spectra.
For each event the integral omnidirectional event-integrated fluences of particles with energy above 30 MeV ($F_{30}$) and above 200 MeV ($F_{200}$)
  were computed using the reconstructed spectra, and the ratio between the two fluences was considered as a simple index of
  the event's hardness.
We also provided a justification of the spectrum estimate in the form of the Band-function, using direct PAMELA data for GLE 71
 (17-May-2012).
We found that, while there is no clear relation between the intensity and the hardness for weak events,
 all strong events with the intensity greater 100 \%*hr are characterized by a very hard spectrum.
This implies that a hard spectrum can be securely assumed for all extreme GLE events, e.g., those studied using cosmogenic
 isotope data in the past.
\end{abstract}


\end{frontmatter}

\parindent=0.5 cm

%%%%%%%%%%%%%%%%%%%%%%%%%%%%%%%%%%%%%%%%%%%%%%%%%%%%%%%%%%%%%%%%%%%%%%%%%%%%%
%% Main text
\section{Introduction}

Strong energy releases may sporadically occur on the Sun, leading to transient phenomena in the interplanetary space.
In particular, solar energetic particle (SEP) events may take place with short but very intensive (by orders of
 magnitude) increases of the flux of energetic particles.
Such events are caused by high fluxes of solar energetic particles (SEPs) accelerated in the solar corona
 and interplanetary space by shocks driven by coronal mass ejections, and by solar flares
 to energies sufficiently high to be able to penetrate into the Earth's atmosphere,
 where they initiate atmospheric cascades whose nucleonic component can be registered by ground-level
 detectors \citep[e.g.,][]{shea_SSR_00,andriopoulou11}, that is called a Ground Level Enhancement (GLE).
Such events were firstly identified using ground-based ionization chambers \citep{forbush_three_1946} and since the 1950s they are
 monitored by the neutron monitor (NM) network.
The first and the strongest GLE detected by the NM network took place on 23-Feb-1956 and recorded as GLE number 5.
GLE events are numbered consequently since then.
The most recent officially accepted GLE was on 17-May-2012, numbered as GLE 71.
Several weak events \citep{thakur14,belov15} have been detected by a few polar NMs even after that date, called sub-GLE events.
However, since they were very weak they are not of interest for this study.
All the GLEs, starting from number 5 are archived at the International GLE database http://gle.oulu.fi \citep{usoskin_GLE_15}.

A NM is an energy-integrated device which cannot measure the differential energy spectrum of primary cosmic-ray particles.
However, for many applications it is important to know the spectrum.
The use of the world-wide NM network makes it possible to assess the integral spectrum, but still an assumption on the spectra shape is needed.
First descriptions were based on the assumption that a GLE spectrum can be described by an exponential over rigidity
 \citep{freier_radiation_1963} or a power law with an exponential roll-off \citep{elisson85}.
However, these simple approximations often do not work well, especially for high energies above several GeV \citep{shea_space_2012}.
As an alternative, the Band-function \citep{band_batse_1993} was proposed as a suitable model to parameterize the
 event-integrated fluence \citep{tylka09}.
The Band-function describes the integral rigidity spectrum by a double power law in rigidity with a smooth
 exponential junction inbetween \citep{usoskin_ACP_11}.
This approximation describes the integral spectrum by a double power law in rigidity with a smooth roll over in-between.
A tremendous work has been performed by \citet{tylka_survey_2008, tylka09} to make a Band-function
 fit to almost all (59) GLE events, using both NM data for high-energy tail and in-situ space-borne measurements
 for the lower part of the spectrum.
We based our present analysis on the result of this extensive work, updated recently (Allan Tylka, personal communication 2015).

The most distinctive feature of GLE events is the hardness of their energy/rigidity spectra \citep{shea_SSR_00}.
As a measure of the intensity of a SEP event, the event-integrated fluence of SEPs with energy above 30 MeV is often used.
Although it is intuitively expected that all GLE events should have a hard spectrum, that is not true.
For example, GLE 24 (Aug 1972) was moderately strong but it provided the largest fluence (omnidirectional flux integrated over the
 total duration of the event) of SEPs with energy above 30 MeV (called henceforth $F_{30}$), greater than
 that of the strongest GLE 5 in February 1956 \citep{belov05}.
The former event had a very soft spectrum, while the latter -- a very hard one.
All other GLEs have a wide variety of spectra between these two cases.
We note that, while ground based NMs are sensitive to relatively high energy part of the SEP spectrum,
 for many application in atmospheric sciences, climate, dosimetry, etc., it is important to know the fluence of lower
 energy particles, with energy above 30 MeV, $F_{30}$.
In particular, strong SEP events can lead to essential changes in the polar atmosphere and even affect
 regional climate \citep[see, e.g.,][]{mironova15}.

Over the last decades, the lower-energy part of the SEP spectrum, which cannot be assessed by NMs, was evaluated from satellite-borne data
 \citep{vainio09,bazilevskaya14}.
On the other hand, when one goes back in time, strong SEP/GLE events can be estimated from indirect proxies -- cosmogenic
 radionuclides produced by cosmic rays in the atmosphere and stored in natural archives, such as
 tree trunks or ice cores \citep{beer12}.
Looking for spikes in cosmogenic data, one may find extreme SEP events in the past
 \citep[e.g.,][]{miyake12,usoskin_ApJ_12}.
It has been recently estimated that production of cosmogenic nuclides can be used as a measure of
 SEP fluence with energy above 200 MeV, called $F_{200}$ \citep{usoskin_F200_14}.
However, while cosmogenic proxy are somewhat more sensitive to SEP comparing to  the ground-based NMs, they are still
 incapable to evaluate the low-energy range of the SEP spectrum, which is most important for atmospheric processes,
 viz. $F_{30}$.
Accordingly, it would be useful to know wether GLE data can provide at least a first order estimate for the
 low-energy fluence of SEP when the higher-energy fluence $F_{200}$ is known.

Different aspects of the GLE event statistic have been studied earlier \citep[see, e.g.][]{cliver06,belov10}.
Here we perform a statistical analysis searching for a relation between the event-integrated intensity
 of GLE events and the hardness
 of their spectra, using the full database of GLEs for the last 60 years.

\section{Analysis} \label{Sec1}

\subsection{GLE strength}

\begin{figure}[h]
\centering \resizebox{\textwidth}{!}{\includegraphics{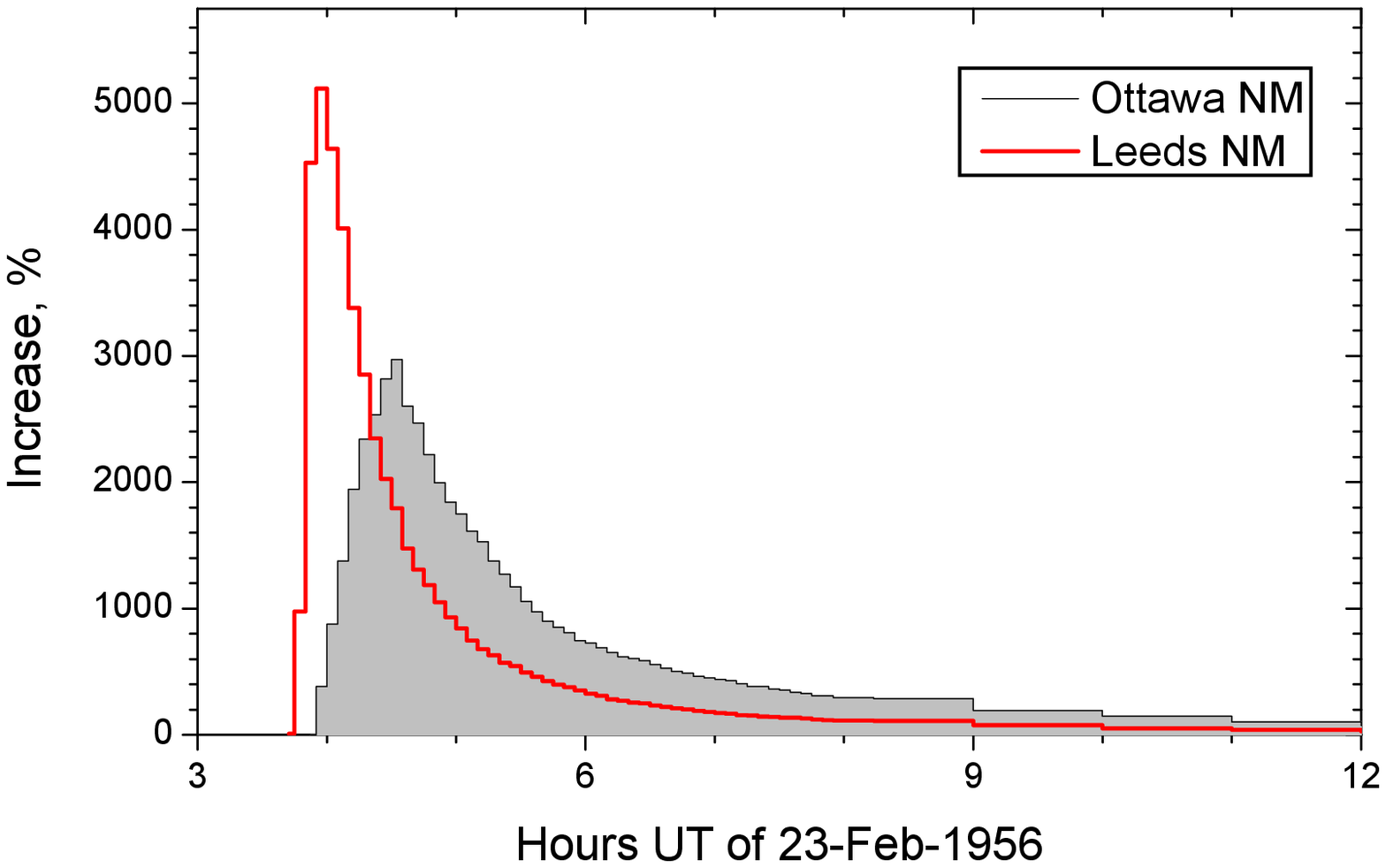}}
\caption{ Percentage increase measured by two NMs, Leeds and Ottawa, during the GLE 5.
While Leeds NM registered the highest peak increase (5116 \% in pseudo 5-min data), Ottawa NM
 detected the most intense integral response (5300 \%*hr).
}
\label{Fig:GLE5}
\end{figure}

It is common to characterize the strength of a GLE as the peak intensity in percentage of the increase above
 the GCR background \citep[e.g.,][]{andriopoulou11}.
For example, the event of 23-Feb-1956 (GLE 5) was characterized by the highest 5116 \% increase in
 pseudo\footnote{The 5-min resolution data for the Leeds NM were interpolated from a graph of the original data
  with 15-min resolution (E. Eroshenko, personal communication 2016).}
 5-min data of the Leeds NM.
However, for some purposes it is more useful to study the integral intensity of the events, so that
 the same GLE 5 had the largest increase of $\approx$5300 \%*hr in Ottawa NM, while it was 4450 $\%$*hr
 in Leeds station (see Fig.~\ref{Fig:GLE5}).
The event-integrated intensity $I$ is defined as the integral of the excess above the GCR background over the entire duration of
 the event (see the shaded area in Fig.~\ref{Fig:GLE5}) and is given here in units of \%*hr.
It corresponds to the total fluence of SEPs with energy sufficient to cause an atmospheric cascade (several hundred MeV).
We note that, while the peak intensity is important for the problem related to particle acceleration and transport
 in the interplanetary medium, the total fluence is more relevant for the terrestrial effect \citep{usoskin_ACP_11}.
Moreover, the integral intensity is much more robustly defined than the peak intensity for the following reasons \citep[e.g.,][]{dorman04}:
 the peak intensity may be distorted for greatest events by the dead-time of the NM (for example, a NM
 with long dead-time of 1.2 ms can lost $\approx 40$\% of counts for a 50-fold increase of the counting rate),
 but this is small for the event-integrated intensity;
 the peak intensity depends on the time resolution, while the intensity is non-dependent on it;
 the peak intensity, especially at the impulsive phase of the event may depend on exact location of NM.

The values of $I$ were computed for GLEs 5 throughout 71 using data collected at the International GLE database
 http://gle.oulu.fi.
The relative increase of GLE vs. the GCR background depends on the local geomagnetic rigidity cutoff
 and the altitude of the NM location.
Here we consider as the strength of the GLE, the value of $I$ averaged over 30 polar sea-level NMs, in order to consider
 a homogeneous response of the NM network to each event.
For each event we considered all the NMs with the geomagnetic cutoff rigidity $<1.5$ GV and the altitude less than 200 m
 above the sea level.
The $<1.5$ GV limit is smooth because of the long-term variations of the cutoff rigifity \citep{kudela04b}.
The list of GLEs with their strengths and the number of the used NMs is given in Table~\ref{Tab1}.
Typical values of $I$ vary a few \%*hr to 5300 \%*hr.
Events with the values of $I$ less than 3 \%*hr are not listed.
We arbitrarily divide events into weak ($I<100$ \%*hr) and strong ($I> 100$ \%*hr).

\subsection{Energy spectrum}
\label{S:spec}
Here we used energy/rigidity spectra of each GLE event as reconstructed by
 \citet{tylka09} from both satellite-borne and ground-based NM data and updated using the newest data
 (Allan Tylka, personal communication, 2015).
The omnidirectional event-integrated fluence for each event is parameterized
 using the Band function, see formalism in \citep{tylka09,usoskin_F200_14}.
For each event we computed, using the spectrum reconstructions, the ratio of two fluences,
 $F_{30}=J(>30$ MeV) and $F_{200}=J(>200$ MeV), which is listed in Table \ref{Tab1}.
The error bars of the ratio were estimated from the propagated uncertainties of the Band-function fits \citep{tylka09}.
We note that the highest ratio of $\approx 480$ appears for GLE 24, where the the fit appeared unphysical
 below 10 MeV, giving a spectrum rising in energy and therefore the parameters of the Band-function fitting were
 estimated omitting the 10 MeV point (Allan Tylka, personal communication, 2015).
We also note that the source region for GLE 24 was near the central meridian.
Accordingly, the event might have included a lower-energy energetic storm particle (ESP) event, caused by
 the CME-driven shock's arrival at Earth.
In this case, the streaming-limit does not apply.
For other near-central-meridian events with clear ESP phases (19-Oct-1989, 24-Aug-1998, 14-Jul-2000, 4-Nov-2001, 28-Oct-2003)
 \citet{tylka09} provided separate fits for the ``GLE'' and ``ESP'' phases.
We used the sum of the two components to estimate the total fluence of SEP particles for such events.

\begin{figure}
\centering \resizebox{\textwidth}{!}{\includegraphics{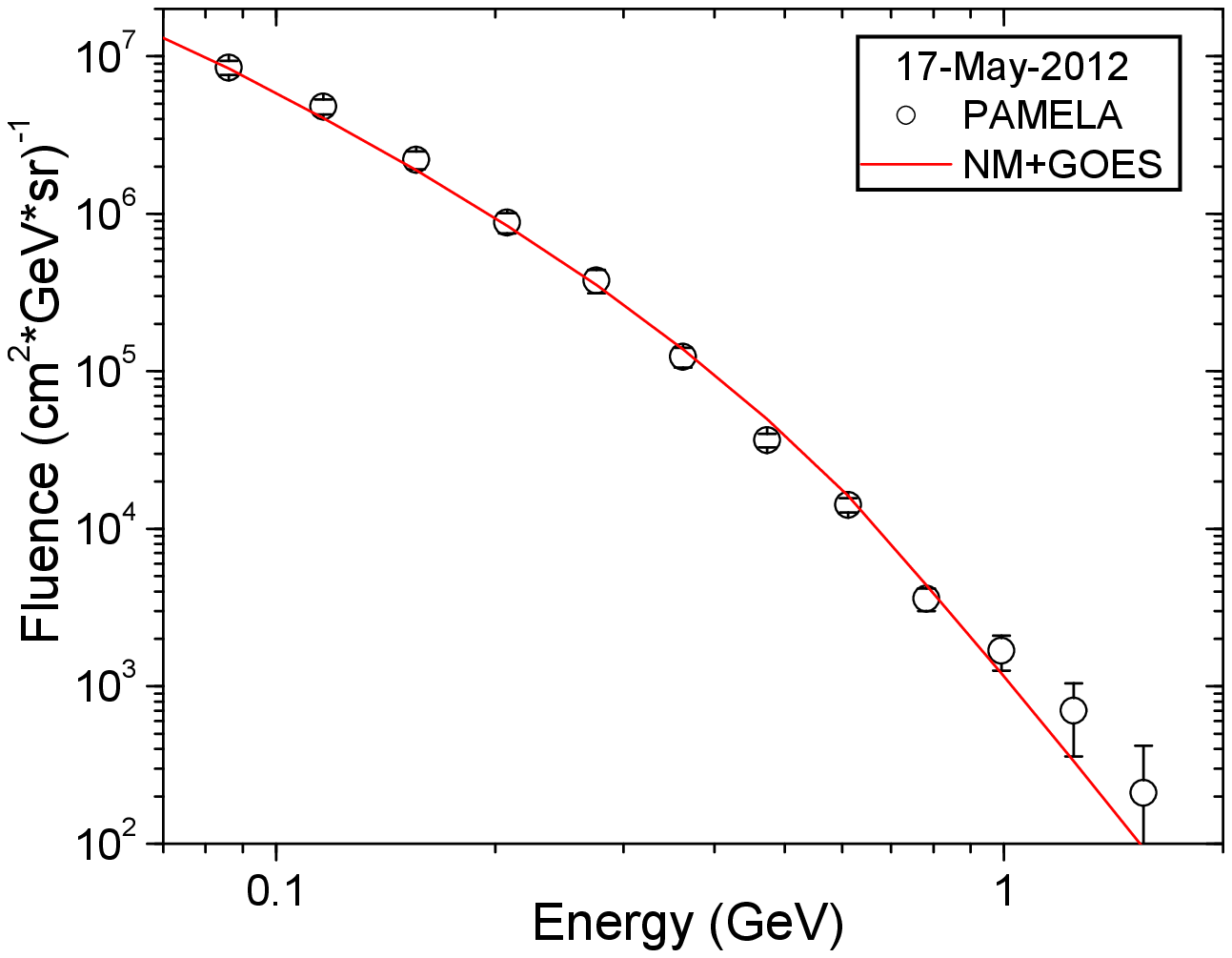}}
\caption{The differential energy fluence of solar energetic particles for the GLE 71 (17-May-2012).
Dots with error bars (only statistical errors are shown, no systematics) represents data (protons only) from the PAMELA experiment
 \citep{bazi_icrc_13}, while the red curve is for the Band-function approximation from \citet{tylka09}.
}
\label{Fig:2012}
\end{figure}

\subsection{GLE 71 with PAMELA data} \label{Sec2}

Here we perform a test of correctness of the Band-function fit to the SEP spectrum using direct space-borne
 measurements.
For some GLEs during the last decade, direct measurements of SEP energy spectra could have been made with space-borne
 magnetic spectrometers such as PAMELA (Payload for Antimatter Matter Exploration and Light-nuclei Astrophysics) mission
 \citep{adriani11}, installed onboard the low orbiting satellite Resurs-DK1 with a quasi-polar (inclination $70^\circ$)
 elliptical orbit (height 350--600 km).
PAMELA is in operation since Summer 2006 and continuously measures all energetic ($>80$ MeV) particles in space.
Thus, it could potentially measure SEP spectra for two GLE events analyzed here, viz. GLE 70 (13-Dec-2006) and 71 (17-May-2012).
No other events can be studied using direct spectral measurements.
If more events are measured they will be analyzed in a due course.
However, as any low orbiting satellite, PAMELA can measure SEP only for a fraction of time, when it is in {the} polar part of its orbit,
 since SEPs are shielded by the geomagnetic field in the lower latitude part of the orbit.

For the GLE 70 PAMELA detected only the initial phase of the event, until about 10 UT of 13-Dec-2006, followed by a
 nearly 24-hr long data gap, making it impossible to study the event-integrated fluence \citep{adriani_apj_11}.

{Measurements of GLE 71 were significantly better}: spectra were measured about $^1\hspace{-0.1cm}/\hspace{-0.05cm}_6$ of the time,
 in the wide energy range with roughly half an hour cadence (polar passes of the orbit) \citep{bazi_icrc_13}.
Thus, the spectrum can be interpolated between the measured points to evaluate the total
 spectral fluence of the SEPs during the event.
The differential energy fluence of the GLE 71, as measured by PAMELA and reconstructed from NMs, is shown in Fig.~\ref{Fig:2012}.
One can see that the two spectra agree quite well in the energy range below 1 GeV, but the errors increase
 for the PAMELA data beyond 1 GeV mostly because of the uncertain contribution of GCR.
We note that PAMELA data were not used in the construction of the Band-function fit for this event.
Unfortunately, the PAMELA spectrum does not go down to 30 MeV, and we cannot obtain the $F_{30}$ fluence directly from
 this data, but considering the lower bound of the energy range for PAMELA data (about 80 MeV), we can assess the ratio of
 $F_{80}/F_{200}$.
It appears $6.15\pm 0.9$ for direct PAMELA data and $6.02\pm 1.38$ for the parameterized spectrum.

Accordingly, we conclude that the Band-function parametrization used here is in good agreement with direct measurements
 of SEPs for the only event, GLE 71, where such a comparison is possible.

\section{Discussion}

\begin{figure}
\centering \resizebox{\columnwidth}{!}{\includegraphics{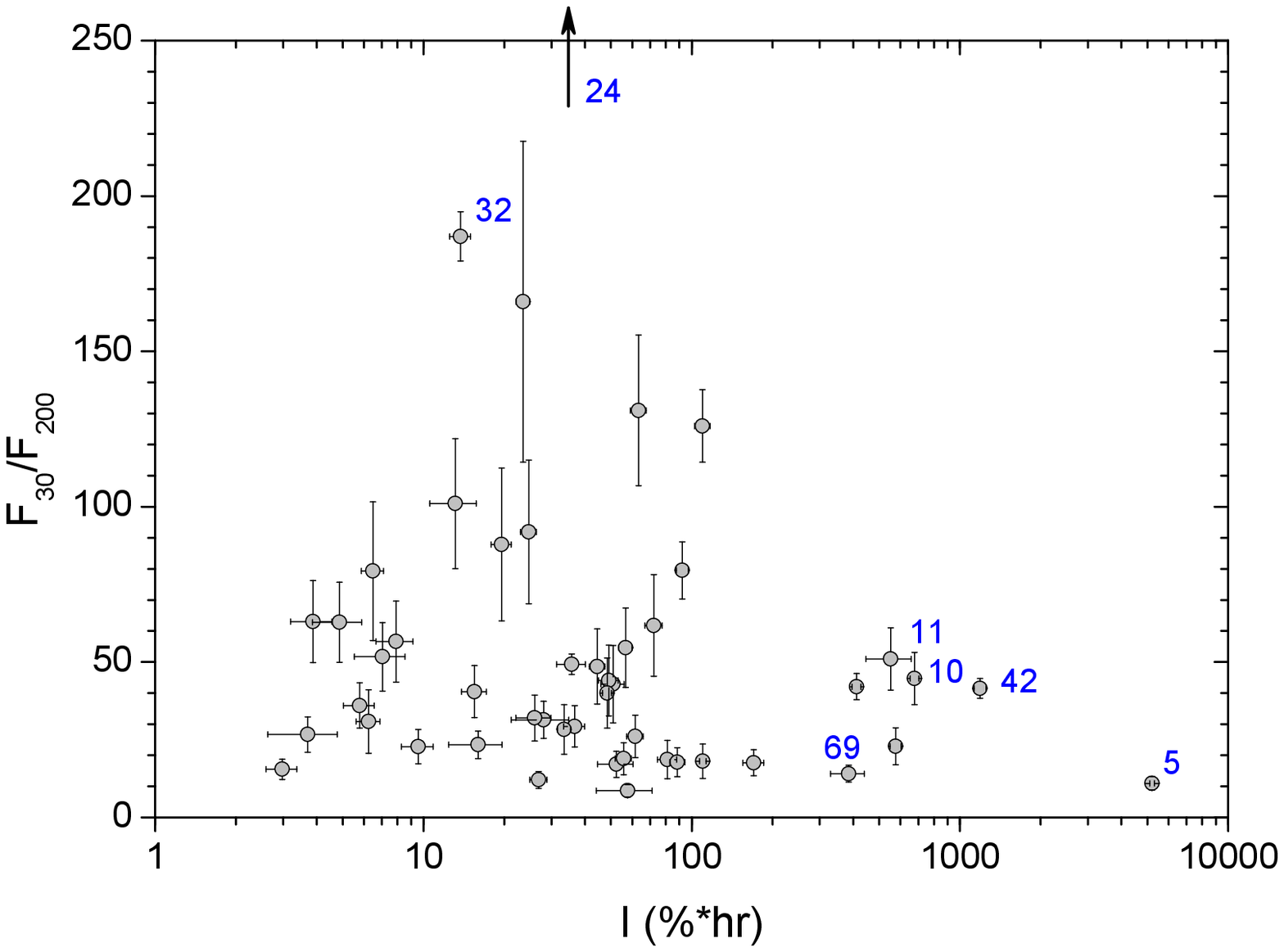}}
\caption{Ratio of fluences $F_{30}$ and $F_{200}$ as a function of the GLE strength $I$.
{GLE numbers (see Table~\ref{Tab1}) are given as blue labels for some points.}
The upward pointing arrow indicates the point for GLE 24 that lies beyond the plot margin.
}
\label{Fig:f_per}
\end{figure}

We show in Figure \ref{Fig:f_per} the $F_{30}/F_{200}$ ratio as a function of the GLE strength for the analyzed events.
One can see that for weak GLE events ($I<100$ \%*hr) the ratio takes a wide range of values, from 10 to 200
 (even more for the GLE 24).
This implies that these GLEs can be with different hardness of the spectrum -- from very hard to very soft.
Interestingly, very weak GLEs ($I<10$ \%*hr) are harder than moderate events ($10<I<100$ \%*hr).
On the other hand, all strong events ($I>100$ \%*hr) are characterized by a hard spectrum -- the ratio
 is limited {to the range} 20--50.
The greatest GLE 5 {has a ratio of about ten, implying a very hard spectrum}.

\begin{figure}
\centering \resizebox{\columnwidth}{!}{\includegraphics{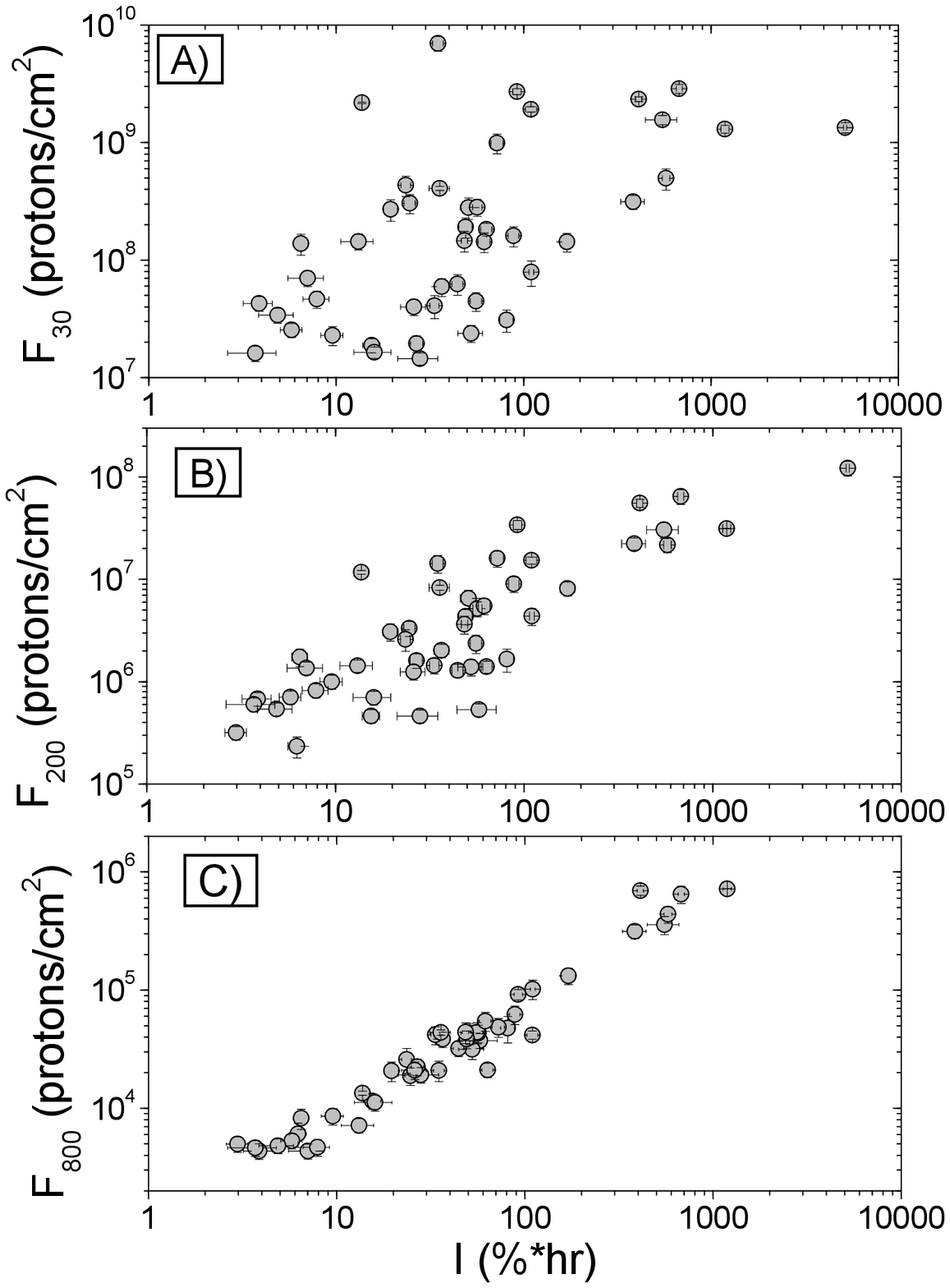}}
\caption{Fluences {$F_{30}$, $F_{200}$ and $F_{800}$ (panels A through C, respectively) as a function of the GLE strength $I$.}}
\label{fig:both}
\end{figure}

Figure \ref{fig:both} depicts the two fluences as function of the GLE strength.
The $F_{30}$ fluence (panel A), while increasing with $I$, shows only insignificant correlation
 (correlation coefficient $0.16\pm0.15$).
Interestingly, except for GLE 24 (see section \ref{S:spec}),
 the $F_{30}$ fluence does not exceed the value of 3$\cdot 10^9$ particles/cm$^2$.
This may be related to a saturation, e.g., the streaming limit \citep{reames10,reames13} caused by a possible
 resonance interaction between the particle flux and the plasma waves at the interplanetary shock so
 that there is a limit for the flux of SEPs accelerated at one instance.
We note that our empirical limit is close to a realistic maximum $F_{30}$ fluence related to the streaming limit,
 estimated by \citet{mccracken01} as (6--8)$\cdot 10^9$ cm$^{-2}$.

\begin{figure}
\centering \resizebox{\columnwidth}{!}{\includegraphics{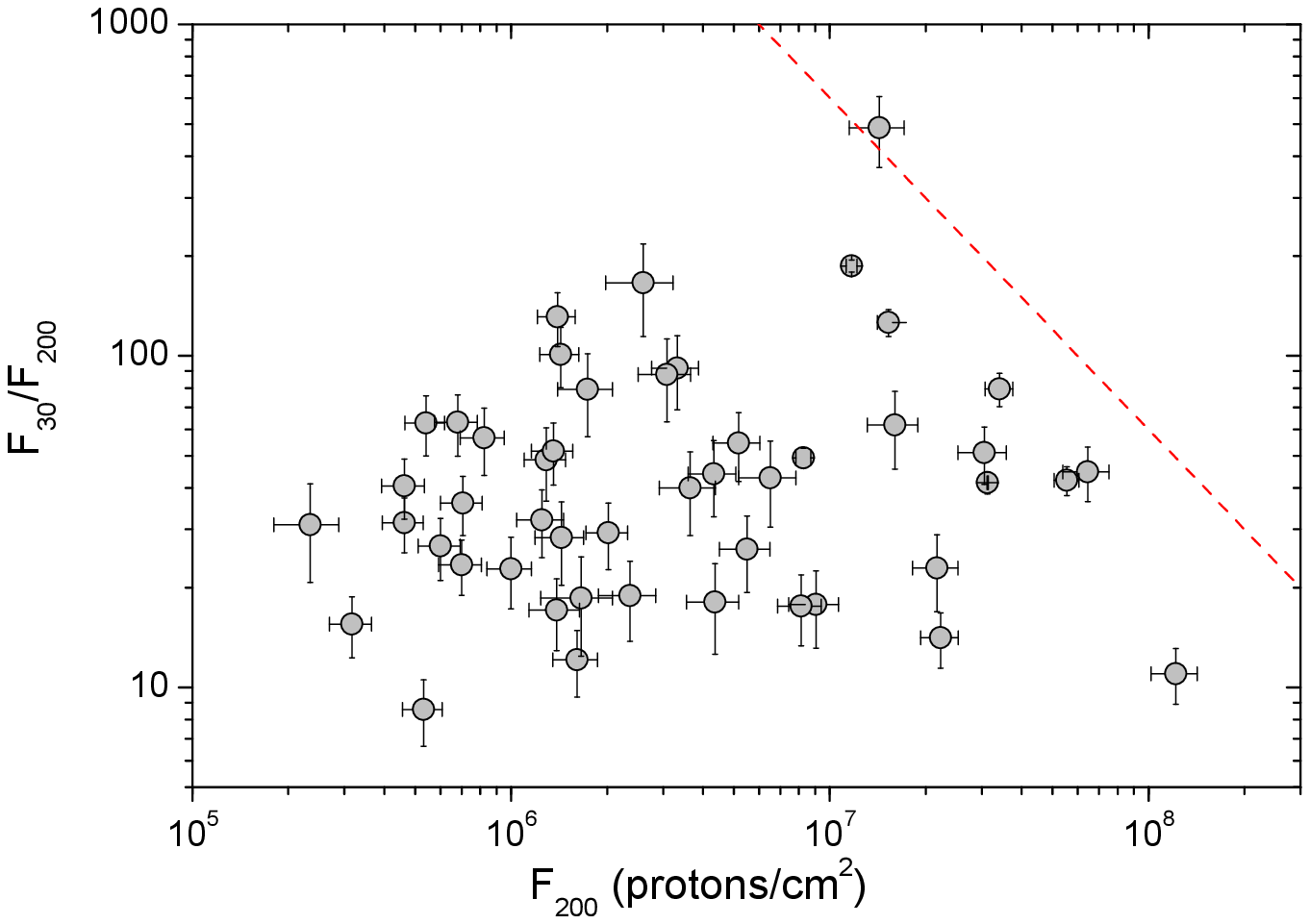}}
\caption{Ratio of fluences $F_{30}$ and $F_{200}$ as a function of the $F_{200}$ fluence.
The red dotted line depicts the ratio expected for the streaming limit of $F_{30}=6\cdot 10^9$ cm$^{-2}$. }
\label{fig:R_F200}
\end{figure}

On the other hand, the $F_{200}$ fluence (panel B) varies almost linearly with the value of $I$, {having a Pearson's
 linear correlation coefficient to be $0.87^{+0.03}_{-0.04}$.}
We note that the best correlation ($0.988^{+0.003}_{-0.004}$) is obtained between the GLE strength $I$ and fluence $F_{800}$
 {(panel C)}, viz. above 800 MeV.
The $F_{200}$ fluence shows no sign of saturation, probably, because the 200 MeV protons do not reach the streaming limit
 because of the falling spectrum.
Accordingly, this may lead to `hardening' of the SEP spectrum for strong events so that $F_{200}$ {continues to increase} with the
 event strength, while $F_{30}$ is saturated at the level of (6--8)$\cdot 10^9$ particles per cm$^{2}$.
The expected in this case ratio $6\cdot 10^9 /F_{200}$ is shown in Figure \ref{fig:R_F200} as the red dotted line.
One can see that it agrees with the observed `hardening' of the spectrum.
Although such a saturation was not directly observed for the $F_{200}$ fluence, it may likely reach
 the streaming limit for extreme events, such as the one of 775 AD, which is
 recognized as the strongest SEP event over the last millennia \citep{usoskin_ApJ_12,mekhaldi15}.

Thus, for {the} strongest events, the higher-energy tail of the spectrum may increase while the lower-energy part of the spectrum is saturated,
 leading to the observed hardening of the spectrum.

\section{Summary}

We studied here the energy spectra of 59 out of 71 GLEs for which there are reconstructions of the event-integrated spectra
 in the form of Band-functions in a wide energy range.
{Using the direct space-borne PAMELA data for GLE 71 (17-May-2012) we confirmed in Section~\ref{S:spec}
 the use of the Band-function reconstructions.}
As the strength of each GLE event we considered the event-integrated response of polar sea-level NMs to SEP, above the
 GCR background.
As a measure of the hardness of the energy spectrum we considered the ratio of $F_{30}/F_{200}$.

We have shown that all the strong GLE events (the strength $I>100$ \%*hr) are characterized by a hard spectrum with the
 $F_{30}/F_{200}$ ratio being 10--50.
Moreover, there seems to be a saturation, most likely due to the streaming limit, of the $F_{30}$ fluence at the level
 of (6--8)$\cdot 10^9$ particle/cm$^2$ so that it does not exceed this level with increasing the strength of the event.
This implies that strongest GLE-like events have very hard spectrum {in the energy range below several hundred MeV},
 and that we do not expect to see an extreme
 GLE event with a soft spectrum in the energy range of tens-hundreds MeV.

This has a practical application for historical SEP events studied using cosmogenic isotope records \citep[e.g.,][]{usoskin_ApJ_12}.
As discussed by \citet{usoskin_F200_14}, cosmogenic isotope data allows reconstructions of the $F_{200}$ while a lower-energy
 range (several tens of MeV) is important for terrestrial effects (ionization of the polar atmosphere or radiation hazards).
Accordingly, it is a reasonable assumption that all strong SEP events identified in cosmogenic isotope data, such as
 the famous event of 775 AD, which was the strongest know over ten millennia \citep{miyake12,usoskin_775_13} have very
 hard spectra \citep[cf., ][]{mekhaldi15}.

\section*{Acknowledgements}
This work was supported by the Center of Excellence ReSoLVE (project No. 272157).

\begin{table}
\caption{Parameters of the studied GLE events.
The columns provide: GLE number and date, the GLE strength (event-integrated intensity), the ratio of $F_{30}/F_{200}$, and the number
 of polar sea-level NMs used to calculate the strength $I$ (see text for details).}
\label{Tab1}
\tiny
\begin{tabular}{lc|ccc}
\hline
GLE & Date & I (\%*hr) & $F_{30}/F_{200}$ & N\\
\hline
5&23-Feb-1956 &5202$\pm$104&11.0$\pm$2.1&2\\
8&04-May-1960&58$\pm$14&8.6$\pm$1.9&9\\
10&12-Nov-1960&677$\pm$25&44.7$\pm$8.4&14\\
11&15-Nov-1960&552$\pm$106&51.0$\pm$10.0&15\\
13&18-Jul-1961&51$\pm$5&42.9$\pm$12.5&13\\
16&28-Jan-1967&110$\pm$3&18.1$\pm$5.5&19\\
19&18-Nov-1968&6$\pm$1&79.3$\pm$22.3&15\\
21&30-Mar-1969&81$\pm$7&18.6$\pm$6.2&18\\
22&24-Jan-1971&25$\pm$2&91.9$\pm$23.1&20\\
23&01-Sep-1971&88$\pm$6&17.8$\pm$4.7&18\\
24&04-Aug-1972&35$\pm$2&488.1$\pm$118.2&16\\
25&07-Aug-1972&20$\pm$2&87.8$\pm$24.6&17\\
26&29-Apr-1973&6$\pm$1&30.9$\pm$10.2&12\\
27&30-Apr-1976&6$\pm$1&36.0$\pm$7.3&13\\
28&19-Sep-1977&4$\pm$1&63.0$\pm$13.2&9\\
29&24-Sep-1977&52$\pm$8&17.1$\pm$4.2&11\\
30&22-Nov-1977&37$\pm$3&29.3$\pm$6.7&12\\
31&07-May-1978&28$\pm$7&31.4$\pm$5.9&12\\
32&23-Sep-1978&24$\pm$1&166.3$\pm$51.6&13\\
36&12-Oct-1981&63$\pm$4&130.9$\pm$24.2&11\\
37&26-Nov-1982&16$\pm$2&40.5$\pm$8.4&11\\
38&08-Dec-1982&44$\pm$3&48.6$\pm$12.1&12\\
39&16-Feb-1984&16$\pm$4&23.4$\pm$4.4&13\\
41&16-Aug-1989&57$\pm$3&54.6$\pm$12.8&12\\
42&29-Sep-1989&1189$\pm$60&41.5$\pm$3.2&15\\
43&19-Oct-1989&411$\pm$15&42.1$\pm$4.2&13\\
44&22-Oct-1989&72$\pm$5&61.8$\pm$16.3&13\\
45&24-Oct-1989&576$\pm$27&22.9$\pm$6.0&13\\
46&15-Nov-1989&3$\pm$0&15.5$\pm$3.2&5\\
47&21-May-1990&33$\pm$2&28.3$\pm$8.0&14\\
48&24-May-1990&56$\pm$4&18.9$\pm$5.1&14\\
49&26-May-1990&27$\pm$2&12.1$\pm$2.7&13\\
52&15-Jun-1991&49$\pm$4&44.1$\pm$11.4&11\\
53&25-Jun-1992&5$\pm$1&62.8$\pm$12.9&10\\
55&06-Nov-1997&48$\pm$2&40.0$\pm$11.3&9\\
56&02-May-1998&4$\pm$1&26.7$\pm$5.7&4\\
59&14-Jul-2000&92$\pm$5&79.5$\pm$9.2&11\\
60&15-Apr-2001&170$\pm$15&17.6$\pm$4.2&13\\
61&18-Apr-2001&26$\pm$4&32.0$\pm$7.4&13\\
62&04-Nov-2001&14$\pm$1&186.6$\pm$7.9&11\\
63&26-Dec-2001&7$\pm$2&51.7$\pm$11.0&11\\
64&24-Aug-2002&8$\pm$1&56.6$\pm$13.0&11\\
65&28-Oct-2003&110$\pm$7&125.6$\pm$11.7&14\\
66&29-Oct-2003&36$\pm$4&49.3$\pm$3.3&14\\
67&02-Nov-2003&13$\pm$3&100.6$\pm$20.9&11\\
69&20-Jan-2005&385$\pm$55&14.1$\pm$2.7&14\\
70&13-Dec-2006&62$\pm$4&26.1$\pm$6.8&15\\
71&17-May-2012&10$\pm$1&22.8$\pm$5.5&11\\
\hline
\end{tabular}
\end{table}

%%%%%%%%%%%%%%%%%%%%%%%%%%%%%%%%%%%%%%%%%%%%%%%%%%%%%%%%%%%%%%%%%%%%%%%%%%%%%
%% Appendices
% \appendix

%\bibliographystyle{elsarticle-harv}      % basic style, author-year citations
%\bibliography{GLE_bib}

\end{document}